\documentstyle[12pt,epsf,aps]{revtex}

\newcommand{\be}{\begin{equation}}
\newcommand{\ee}{\end{equation}}
\newcommand{\bea}{\begin{eqnarray}}
\newcommand{\eea}{\end{eqnarray}}

\begin{document}


\title
{
 Vacuum polarization induced coupling between Maxwell
 and Kalb-Ramond fields
 \vspace*{5mm}
}
\author
{
 N. D. Hari Dass\footnote{email : dass@imsc.ernet.in}
 and
 K. V. Shajesh\footnote{email : kvshajesh@yahoo.com}
}
\address
{
 Institute of Mathematical Sciences,
 C. I. T. Campus, Taramani, Chennai - 600 113, India.
}

\maketitle
\vspace{10mm}


\begin{abstract}

We present here a manifestly gauge invariant calculation 
of vacuum polarization to fermions in the presence of
a constant Maxwell and a constant Kalb-Ramond field in
four dimensions.
The formalism is a generalisation of the one used by Schwinger
in his famous paper on gauge invariance and vacuum 
polarization. We get an explicit expression for the 
vacuum polarization induced effective Lagrangian 
for a constant Maxwell field interacting with a 
constant Kalb-Ramond field.
In the weak field limit we get the coupling between the
Maxwell field and the Kalb-Ramond field to be 
$(\tilde{H}.\tilde{F})^2$,
where ${\tilde H}_{\mu}=
{1\over {3!}}\epsilon_{\mu\alpha\beta\lambda}H^{\alpha\beta\lambda}$
and $\tilde F$ is the dual of $F_{\mu\nu}$.

\end{abstract}


\section{Introduction}
{\label{intro}}

Consider the interaction involving the fermion field,
the Maxwell field and the Kalb-Ramond field.
Let us approximate the Maxwell field and the Kalb-Ramond
field to be classical background fields. 
We address to the following question:
what will be the induced effective Lagrangian for the classical
background (consisting of the Maxwell field and the Kalb-Ramond
field) because of the vacuum polarization 
of the quantised fermion field?
This task is achieved by integrating out the fermion field.

The evaluation of the vacuum polarization induced effective Lagrangian 
for a classical background field 
interacting with a quantum field
originated in the works of
W. Heisenberg and H. Euler\cite{heisen} 
and V. Weisskopf\cite{weiss} before the advent of QED.
An explicit expression within QED for the vacuum polarization induced
effective Lagrangian for the classical background electromagnetic
field interacting with the quantised fermion field was 
evaluated by Julian Schwinger in \cite{sch1951}. 

Quantum electrodynamics is a $U(1)$ gauge theory
in four dimensional Minkowski space-time.
It involves a 1-form gauge potential $A_{\mu} (x)$ and a 
matter field $\psi (x)$ acting as the source of $A_{\mu} (x)$.
In QED the source of $A_{\mu} (x)$ is a point particle.
The Lagrangian for the 1-form quantum electrodynamics
\be
{\cal {L}} 
=
- \frac{1}{4} F_{\mu \nu} F^{\mu \nu}
+ \overline{\psi} (x) 
    \left(
      i \not{\partial} - e \not{A} 
      - m
    \right)
  \psi (x)
\ee
is constructed by demanding invariance of the Lagrangian 
under the $U(1)$ gauge transformations
\bea
A_{\mu} (x) 
&\rightarrow&
A_{\mu} (x) 
+ \partial_{\mu} \Lambda^A (x)
\nonumber 
\\
\psi (x) 
&\rightarrow&
\exp \left\{ i e \Lambda^A (x) \right\} \psi (x)
\eea
where $\Lambda^A (x)$ is an arbitrary 0-form field.
$F_{\mu \nu} (x) = \partial_{\mu} A_{\nu} - \partial_{\nu} A_{\mu}$
is the 2-form antisymmetric field tensor constructed from the 1-form
gauge potential $A_{\mu} (x)$.Requirements of renormalisability pose
additional requirements and eqn (1) is a minimal choice.

The p-form quantum electrodynamics 
\cite{teitel,teitel_henn,nambu1976}
is a $U(1)$ gauge theory
constructed out of the p-form gauge potential
$A_{\mu_1 ... \mu_p}$. As the gauge transformation parameter in such
cases is a $p-1$-form, the analog of eqn(2)  makes sense only if $\psi$
represents a $p-1$-dimensional extended object. 
In particular 2-form quantum electrodynamics is a 
$U(1)$ gauge theory 
which involves a 2-form gauge potential $B_{\mu \nu} (x)$
and a string matter field $\psi [ x(\sigma) ]$ for the source
of $B_{\mu \nu} (x)$. 
The Lagrangian for 2-form quantum electrodynamics
\be
{\cal {L}}
=
- \frac{1}{12} H_{\mu \nu \lambda} H^{\mu \nu \lambda}
+ \overline{\psi} [x(\sigma)]~
    \Gamma_{\mu \nu}
    \left(
      i \frac{\delta}{\delta \Omega_{\mu \nu}}
      - \mu A^{\mu \nu}
    \right)
  \psi [x(\sigma)]
\ee
is constructed by demanding invariance 
under the $U(1)$ gauge transformations
\bea
B_{\mu \nu} (x)
&\rightarrow&
B_{\mu \nu} (x)
+ \partial_{\mu} \Lambda^B_{\nu} (x) 
- \partial_{\nu} \Lambda^B_{\mu} (x)
\nonumber
\\
\psi [x(\sigma)] 
&\rightarrow&
\exp \left\{ i \mu \int_{\sigma} 
       \Lambda_{\mu}^B (x) ~dx^{\mu} 
     \right\} 
\psi [x(\sigma)] 
\eea
where $\Lambda_{\mu}^B (x)$ is an arbitrary 1-form field.
$
H_{\mu \nu \lambda} (x)
=
\partial_{\mu} B_{\nu \lambda}
+ \partial_{\nu} B_{\lambda \mu}
+ \partial_{\lambda} B_{\mu \nu }
$
is the 3-form totally antisymmetric field tensor constructed from the
2-form antisymmetric gauge potential $B_{\mu \nu } (x)$.
$H_{\mu \nu \lambda} (x)$ is popularly called the 
Kalb-Ramond field.
The functional derivative 
$\frac{\delta}{\delta \Omega_{\mu \nu}}$
captures the variation of $\psi [x(\sigma)]$ due to an infinitesimal
change of the string configuration which sweeps a surface 
element $d \Omega_{\mu \nu}$.

Kalb-Ramond fields play important roles in string theories and field
theories
\cite{kalb_ramond1974,green_sch_witten}.
In $d=4$ the Kalb-Ramond field is dual to the derivative of a spinless
field $\phi$ which can be interpreted as an axion. We shall not
elaborate too much on the importance of the Kalb-Ramond field here. We
believe that the exercise we have carried out in this paper is
interesting in its own right.

As the system represented by eqns(3,4) is too complicated since it involves string
theory,in this paper we have investigated the simpler problem of the interaction
of a point-like fermion field with Maxwell and 
Kalb-Ramond fields in four dimensional Minkowski space-time.
This is only possible if $\psi$ carries no Kalb-Ramond charge.
Nevertheless, we show that a nontrivial effective Lagrangian between
Maxwell and Kalb-Ramond fields emerges.
The simplest Lagrangian we can construct by demanding 
invariance under the $U(1)$
gauge transformations
\bea
B_{\mu \nu} (x)
&\rightarrow&
B_{\mu \nu} (x)
+ \partial_{\mu} \Lambda^B_{\nu} (x) 
- \partial_{\nu} \Lambda^B_{\mu} (x)
\nonumber
\\
A_{\mu} (x) 
&\rightarrow&
A_{\mu} (x) 
+ \partial_{\mu} \Lambda^A (x)
\nonumber 
\\
\psi (x) 
&\rightarrow&
\exp \left\{ i e \Lambda^A (x) \right\} \psi (x)
\eea
is
\be
{\cal {L}}
=
- \frac{1}{4} F_{\mu \nu} F^{\mu \nu}
- \frac{1}{12} H_{\mu \nu \lambda} H^{\mu \nu \lambda}
+ \overline{\psi} 
    \left(
      i \not{\partial} - e \not{A} 
      - \frac{1}{12} \frac{g}{m} 
        \sigma_{\mu \nu \lambda} H^{\mu \nu \lambda}
      - m
    \right)
  \psi
.
\label{eqn1.30}
\ee
Such a Lagrangian could arise 
from string theories after compactification
to four dimensional space-time.

Schwinger's calculation in \cite{sch1951}
involves evaluation of the 
vacuum polarization induced effective Lagrangian
for the case of 1-form quantum electrodynamics.
In this work we generalize this 
to the Lagrangian given in eqn. (\ref{eqn1.30}).
We evaluate an explicit expression for the vacuum polarization
induced effective Lagrangian 
for the classical
background (consisting of Maxwell and Kalb-Ramond
fields) interacting with the quantised fermion field.
We present the paper exactly along the lines of
Schwinger's paper.
As can be seen subsequently, there are important differences between 
our case and that of Schwinger.
Due to these reasons our calculation could give some insight
if one attempts to generalize Schwinger's calculation for other
circumstances like the
non abelian gauge theories.
An important check over our calculations is that at every stage 
we are able to  reproduce Schwinger's results 
for QED by taking the 
limit $g \rightarrow 0$.

In section \ref{formalism} 
we formulate the methodology used by Schwinger 
in \cite{sch1951} for a general background field.
We integrate out the fermion field.
As in \cite{sch1951} the determination of the vacuum polarization
induced effective Lagrangian reduces to the evaluation of
the trace of the exponential of a suitable operator.
Schwinger 
interprets the exponential of the operator 
as an evolution operator describing the dynamics of
a ``particle'' with a ``Hamiltonian'' evolving in ``proper time''.

In section \ref{constant} we exactly solve the 
evolution operator for the case of a constant Maxwell field 
and a constant Kalb-Ramond field. 
The Hamiltonian under consideration
can be understood as a generalization of 
Schwinger's case.
There are additional 
complications in our case arising out of operators not commuting with each other.

The effective Lagrangian
is determined by evaluating the trace of the evolution operator
in both the coordinate index and the spinor index.
This involves solving for the eigenvalues of the 
Maxwell field $F_{\mu \nu}$ in the coordinate index
and of another operator $Q$ in the spinor index.
The determination of the eigenvalues is simplified by
constructing suitable polynomial equations
which are satisfied by $F_{\mu \nu}$ and $Q$.
The eigenvalues are conveniently expressed in terms of
gauge and Lorentz invariant quantities. 

We thus get an explicit expression for the vacuum polarization
induced effective Lagrangian for the case of constant
field strengths in terms of gauge invariant quantities.
Renormalization of the two field strengths
$F_{\mu \nu}$ and $H_{\mu \nu \lambda}$, and the two
coupling constants $e$ and $g$ leads to a divergence
free expression for the effective Lagrangian.
For the case of weak field strengths we show that
the leading order interaction between the Maxwell field
and the Kalb-Ramond field is given by
$(\tilde{H} . \tilde{F})^2$,
where ${\tilde H}_{\mu}=
{1\over {3!}}\epsilon_{\mu\alpha\beta\lambda}H^{\alpha\beta\lambda}$
and $\tilde F$ is the dual of $F_{\mu\nu}$.


\section{Formulation of the general problem}
{\label{formalism}}

In this section we formulate Schwinger's techniques 
for the case of a fermion field interacting with  external  
Maxwell and  Kalb-Ramond fields.
In the next section we obtain the exact solutions for the case of  
constant external fields.

In eqn. (\ref{eqn1.30}) we have seen that the
Lagrangian for a spin $\frac{1}{2}$ fermion field 
interacting with a Kalb-Ramond field and a Maxwell field 
is given by
\be
{\cal {L}} [H, F, \psi]
=
- \frac{1}{4} F_{\mu \nu} F^{\mu \nu}
- \frac{1}{12} H_{\mu \nu \lambda} H^{\mu \nu \lambda}
+ \overline{\psi} 
    \left(
      i \not{\partial} - e \not{A} 
      - \frac{1}{12} \frac{g}{m} 
        \sigma_{\mu \nu \lambda} H^{\mu \nu \lambda}
      - m
    \right)
  \psi
,  
\ee
where $\psi (x)$ is the fermion field.
Our aim is to evaluate the effective Lagrangian arising from fermionic
vacuum polarisation at one loop level.
This amounts to 
integrating out the contribution of the fermion field.
Using the above prescription we can evaluate the effective 
Lagrangian ${\cal {L}}_{eff} [F, H]$
from the expression
\bea
e^{i \int d^4 x~ {\cal {L}}_{eff} [F, H]}
&=&
\frac
{
\int {\cal {D}} \psi \int {\cal {D}} \overline{\psi}~
e^{
   i \int d^4 x~
   \left[
   - \frac{1}{4} F_{\mu \nu} F^{\mu \nu}
   - \frac{1}{12} H_{\mu \nu \lambda} H^{\mu \nu \lambda}
   + \overline{\psi}
       \left(
         i \not{\partial} - e \not{A}
         - \frac{1}{12} \frac{g}{m} 
	   \sigma_{\mu \nu \lambda} H^{\mu \nu \lambda}
         - m
       \right)
     \psi
   \right]
  }
}
{
\int {\cal {D}} \psi \int {\cal {D}} \overline{\psi}~
e^{
   i \int d^4 x~
      \overline{\psi} 
      \left( i \not{\partial} - m \right)
     \psi
  }
}
.
\label{eqn2.10}
\eea
This can be thought of as the path integral realisation of Schwinger's
approach.
Using the Gaussian integral formula for the anticommuting fermion field
\bea
\int {\cal {D}} \psi \int {\cal {D}} \overline{\psi}~
e^{- \int d^4 x~ \overline{\psi} M \psi}
&=&
\mbox{const}~~ \mbox{Det} M
\label{eqn2.15}
\eea
in eqn. (\ref{eqn2.10}) we have
\bea
e^{i \int d^4 x~ {\cal {L}}_{eff} [F, H]}
&=&
e^{i \int d^4 x~ 
\left[
   - \frac{1}{4} F^2
   - \frac{1}{12} H^2
\right] } 
\left[
\frac{\mbox{Det}~
     \left(
         i \not{\partial} - e \not{A}
         - \frac{1}{12} \frac{g}{m} 
	   \sigma_{\mu \nu \lambda} H^{\mu \nu \lambda}
         - m
       \right)
     }
     {\mbox{Det}~( i \not{\partial} - m )}
\right]
\label{eqn2.20}
.
\eea
The determinant $\mbox{Det}$ is over both the coordinate and the
spinor index.
The notation we shall use is, Det(or Tr) to denote 
determinant(or trace) over both the 
coordinate and spinor index, and det(or tr) to denote
determinant(or trace) over only the spinor index.
The determinant of an operator $M$ remains 
invariant under a similarity transformation $S$,
i.e. $\mbox{Det}~M = \mbox{Det}~( SMS^{-1} )$.
Choosing $S$ to be the charge conjugation matrix $C$ we can write
\bea
\mbox{Det}~M 
&=&
\left\{ \mbox{Det}~M 
\cdot \mbox{Det}~ (CMC^{-1}) \right\}^{\frac{1}{2}}
=
\left\{ 
\mbox{Det}~ (MCMC^{-1}) \right\}^{\frac{1}{2}}
,
\label{eqn2.21}
\eea
Using eqn. (\ref{eqn2.21}) in eqn. (\ref{eqn2.20}) we get
\bea
e^{i \int d^4 x~ {\cal {L}}_{eff} [F, H]}
&=&
e^{i \int d^4 x~
\left[
   - \frac{1}{4} F^2 - \frac{1}{12} H^2
\right] }
\left[
\frac{\mbox{Det}~{\cal {H}}}{\mbox{Det}~{\cal {H}}_0}
\right]^{\frac{1}{2}}
,
\label{eqn2.35}
\eea
where
\bea
{\cal {H}}
&=&
\left(
         i \not{\partial} - e \not{A}
         - \frac{1}{12} \frac{g}{m} 
	   \sigma_{\mu \nu \lambda} H^{\mu \nu \lambda}
         - m
  \right)
  \left(
         - i \not{\partial} + e \not{A}
         - \frac{1}{12} \frac{g}{m} 
	   \sigma_{\mu \nu \lambda} H^{\mu \nu \lambda}
         - m
  \right)
\label{eqn2.36}
\\
{\cal {H}}_0
&=&
\left( i \not{\partial} - m \right)
  \left( - i \not{\partial} - m \right)
.
\label{eqn2.37}
\eea
We have used the fact that the gamma matrices transform under
the charge conjugation matrix as 
$
C \gamma_{\mu} C^{-1} 
=
- {\gamma_{\mu}}^T
.
$
Further, the identity 
$\mbox{Det}~ M = \exp \left[ \mbox{Tr}~\mbox{ln}~M \right]$
and eqn. (\ref{eqn2.35}) gives
\bea
\int d^4 x~ {\cal {L}}_{eff} [F, H]
&=&
\int d^4 x~
\left[
   - \frac{1}{4} F^2 - \frac{1}{12} H^2
\right]
- \frac{i}{2}
~\mbox{Tr}~
\left[
(\mbox{ln}~{\cal {H}}) - (\mbox{ln}~{\cal {H}}_0)
\right]
.
\label{eqn2.40}
\eea

Using the integral representation
\bea
\mbox{ln}~M
&=&
- \int_0^{\infty}
\frac{ds}{s} e^{- i M s}
\label{eqn2.45}
\eea
in eqn. (\ref{eqn2.40}) we have 
\bea
\int d^4 x~ {\cal {L}}_{eff} [F, H]
&=&
\int d^4 x~
\left[
   - \frac{1}{4} F^2 - \frac{1}{12} H^2
\right]
+ \frac{i}{2} \int_0^{\infty} \frac{ds}{s}~
\left\{
 \left[
   \mbox{Tr}~
   e^{- i{\cal {H}} s}
 \right]
-
 \left[
   \mbox{Tr}~
   e^{- i{\cal {H}}_0 s}
 \right]
\right\}
.
\label{eqn2.50}
\eea
Convergence requires that $m$ be interpreted as $m-i\epsilon$. Eqn(16)
has to be regularised carefully and this will be discussed later.

Let us denote
\be
U(x^{\prime \prime}, x^{\prime};s)
\equiv <x^{\prime\prime}| e^{- i {\cal {H}} s}|x^{\prime}>
.
\ee
Following Schwinger, we shall treat $s$ as a sort of time variable
called "proper time" by him and use a Heisenberg-like picture, in which 
the operators evolve in time.If $A(0)$ is an operator at $s=0$, the proper time
dependence of the opertor is defined by
\be
A(s)  \equiv e^{i{\cal H}s} A(0) 
e^{-i{\cal H}s}
\ee
and
\be
<x^{\prime\prime}(s)| \equiv <x^{\prime\prime}|
e^{-i{\cal H}s}
\ee
It then follows that
\bea
<x^{\prime\prime}(s)|x(s)&=&<x^{\prime\prime}(s)|x^{\prime\prime}\nonumber\\
x(0)|x^{\prime}>&=&x^{\prime}|x^{\prime}>
\eea
Thus the problem of determining the effective Lagrangian 
reduces to the evaluation of $U(x(s), x(0);s) \equiv e^{-i{\cal H}s}$.
$U(x(s), x(0);s)$ can be interpreted as the operator describing
the development of a quantum mechanical system governed by the 
``Hamiltonian'' ${\cal {H}}$ in the ``time'' $s$ from a state
$|x^{\prime}(0)>$ to another state $|x^{\prime\prime}(s)>$. Care
must be taken to distinguish the operator valued $U(x(s),x(0);s)$
from the c-number valued function $U(x^{\prime\prime},x^{\prime};s)$.

It is useful to choose $x(s)$ and $x(0)$ 
as independent variables.
If we were doing classical mechanics, we would have interpreted
$U(x(s), x(0);s)$ as the generating function for the canonical 
transformation
\bea
x_{\mu} (s) &=& x_{\mu} (x(0), p(0),s)
\nonumber \\
p_{\mu} (s) &=& p_{\mu} (x(0), p(0),s)
.
\eea
and solved for $U(x(s), x(0);s)$ using the Hamilton Jacobi
equations \cite{goldstein}.
Equations corresponding to the Hamilton Jacobi equations
for the quantum mechanical case   are
\cite{fock,nambu1950}
\bea
{\cal {H}} U(x(s), x(0);s) 
&=&
i \frac{\partial}{\partial s} U(x(s), x(0);s)
\label{eqn2.58}
\\
\pi_{\mu} (0) U(x^{\prime\prime}, x^{\prime};s)
&=&
\left\{
- i \frac{\partial}{\partial x_{\mu}^{\prime} } 
- e A_{\mu} (x^{\prime})
\right\}
U(x^{\prime\prime}, x^{\prime};s)
\label{eqn2.59}
\\
\pi_{\mu} (s) U(x^{\prime\prime}, x^{\prime};s)
&=&
\left\{
+ i \frac{\partial}{\partial x_{\mu}^{\prime\prime} } 
- e A_{\mu} (x^{\prime\prime})
\right\}
U(x^{\prime\prime}, x^{\prime};s)
\label{eqn2.60}
\eea
with the initial condition
\bea
\lim_{s \rightarrow 0}
U(x^{\prime\prime}, x^{\prime};s)
&=&
\delta^{(4)}
\left( x^{\prime\prime} - x^{\prime} \right)
.
\label{eqn2.61}
\eea
where $\pi_{\mu} (s) = p_{\mu} (s) - e A_{\mu} (s)$. 
Thus evaluation of $U(x^{\prime\prime}, x^{\prime};s)$ will involve solving the
first order differential equations (\ref{eqn2.58}-\ref{eqn2.60})
with the initial condition in eqn. (\ref{eqn2.61})
in conjunction with the Heisenberg equations of motion
for the operators $x_{\mu} (s)$ and $p_{\mu} (s)$
given by
\bea
\frac{d x_{\mu} (s)}{ds}
&=&
- i~ [x_{\mu} (s), {\cal {H}}]
\nonumber 
\\
\frac{d \pi_{\mu} (s)}{ds}
&=&
- i~ [\pi_{\mu} (s), {\cal {H}}]
\label{eqn2.65}
\eea
The basic strategy is to try and express ${\cal H}$ in terms of
$x_{\mu}(s),x_{\mu}(0)$ such that $x_{\mu}(s)$ always appear to the
left and $x_{\mu}(0)$ always on the right.Then all operators can be traded for
c-numbers and one gets a first order differential equation which is in principle
solvable.
It is not always possible to solve for $U(x^{\prime\prime}, x^{\prime};s)$ exactly.
A general series solution for $U(x^{\prime\prime}, x^{\prime};s)$
can be got using the method by DeWitt\cite{dewitt}.
In the next section we shall exactly solve the above
first order differential equations for the case when $H$
and $F$ are constant fields.
In the general case after solving for $U(x^{\prime\prime}, x^{\prime};s)$ using 
eqn. (\ref{eqn2.58}), eqn. (\ref{eqn2.59}) 
and eqn. (\ref{eqn2.60}) 
we can write 
\bea
\mbox{Tr}~
\left[ e^{- i {\cal {H}} s} \right]
&=&
\mbox{Tr}~\left\{ U(x^{\prime\prime}, x^{\prime};s) \right\}
=
~\mbox{tr}~ \int d^4 x~ 
\left\{ U(x^{\prime\prime}, x^{\prime};s) \right\}_
{x^{\prime\prime}=x^{\prime}=x}
.
\label{eqn2.70}
\eea
Using eqn. (\ref{eqn2.70}) in eqn. (\ref{eqn2.50}) we have
\bea
{\cal {L}}_{eff} [F, H]
&=&
- \frac{1}{4} F^2
- \frac{1}{12} H^2
+ \frac{i}{2} ~\mbox{tr}~ \int_0^{\infty} \frac{ds}{s}~
  \left\{ U(x^{\prime\prime}, x^{\prime};s) - 
  U_0 (x^{\prime\prime}, x^{\prime};s) \right\}_
  {x^{\prime\prime}=x^{\prime}=x} 
.
\label{eqn2.80}
\eea


\section{Constant fields}
\label{constant}

In this section we shall evaluate an exact expression for 
the effective Lagrangian given in eqn. (\ref{eqn2.80})
for the case of constant $F$ and $H$ fields \cite{taub}.
This is achieved by solving for $U(x^{\prime\prime}, x^{\prime};s)$
using eqn. (\ref{eqn2.58}), (\ref{eqn2.59}), (\ref{eqn2.60})
and (\ref{eqn2.65}). 
The divergent terms in the effective Lagrangian are absorbed 
by suitable renormalization of the field strengths
$F_{\mu \nu}$ and $H_{\mu \nu \lambda}$ and the couplings
$e$ and $g$.
Making a series expansion of the effective Lagrangian 
for the case of weak fields we get the leading order
interaction between the Maxwell field and the Kalb-Ramond
field to be $(\tilde{H} . \tilde{F})^2$ analogous to the 
Euler-Heisenberg effective
Lagrangian for electromagnetic fields.

For the case of constant $F$ and $H$ fields the expression for 
${\cal {H}}$ in eqn. (\ref{eqn2.36}) simplifies to 
\bea
{\cal {H}}
&=&
- \pi_{\mu} \pi^{\mu}
- N_{\mu} \pi^{\mu}
- \frac{1}{4} N_{\mu} N^{\mu}
+ m^2 + Q
\label{eqn3.10}
\eea
where
\be
\pi_{\mu} = i \partial_{\mu} - e A_{\mu}
,
\hspace{10mm}
N_{\mu} 
=
- i \frac{g}{m} {\tilde{H}}_{\mu} \gamma_5
\hspace{10mm}
\mbox{and}
\hspace{10mm}
Q =
\frac{1}{2} e \sigma_{\mu \nu} F^{\mu \nu}
+ i g {\tilde{H}}_{\mu} \gamma_5 \gamma^{\mu}
.
\label{eqn3.11}
\ee
where ${\tilde{H}}_{\mu}$ is the dual tensor corresponding 
to $H_{\mu \nu \lambda}$ and is given by
${\tilde{H}}_{\mu} = \frac{1}{3!} 
\epsilon_{\mu \alpha \beta \lambda}
H^{\alpha \beta \lambda}$.
It should be noted that $N_{\mu}$ and $Q$
do not commute with each other.This is the additional complication we
referred to earlier.The above simplification assumes
that $B_{\mu \nu} (x)$ and $A_{\mu} (x)$ are classical 
(not quantised) fields.
We also assume that the corresponding tensor fields
$H_{\mu \nu \lambda} (x)$ and $F_{\mu \nu} (x)$ 
are constants in space and time.
The above simplification also
uses the identities,
\be
\sigma_{\alpha \beta \lambda}
=
i \epsilon_{\alpha \beta \lambda \mu} \gamma_5 \gamma^{\mu}
\hspace{10mm}
\mbox{and}
\hspace{10mm}
\left[ \sigma_{\alpha \beta \lambda} , \gamma_{\sigma} \right]
=
2 i \epsilon_{\alpha \beta \lambda \sigma} \gamma_5
.
\ee
For the case of constant $F$ and $H$ fields the Heisenberg equations
of motion for $x(s)$ and $\pi(s)$ in eqn. (\ref{eqn2.65})
simplifies to 
\bea
\frac{d x_{\mu} (s)}{ds}
&=&
2 \pi_{\mu} + N_{\mu}
\nonumber \\
\frac{d \pi_{\mu} (s)}{ds} 
&=&
e F_{\mu \alpha} \left\{ 2 \pi^{\alpha} (s) + N^{\alpha} \right\} 
.
\label{eqn3.20}
\eea
Using eqn. (\ref{eqn3.20}) we can solve for 
$\pi_{\mu} (s)$ and $\pi_{\mu} (0)$ 
in terms of 
$x_{\mu} (s)$ and $x_{\mu} (0)$
to get
\bea
\vec{\pi} (0)
&=&
\left( \frac{e \hat{F}}{e^{2 e \hat{F} s} - 1} \right)
\cdot
\left[ \vec{x} (s) - \vec{x} (0) \right]
- \frac{1}{2} \vec{N}
\nonumber \\
\vec{\pi} (s)
&=&
\left( \frac{e \hat{F}}{1 - e^{- 2 e \hat{F} s}} \right)
\cdot
\left[ \vec{x} (s) - \vec{x} (0) \right]
- \frac{1}{2} \vec{N}
\label{eqn3.30}
\eea
where we have used the symbolic notation,
`$\vec{a}$' (vector) to denote $a_{\mu}$ and
`$\hat{A}$' (matrix) to denote $A_{\mu \nu}$.
Observe that in this notation $a_{\mu} a^{\mu}$ will
be denoted by $\vec{a}^T \cdot \vec{a}$.
It is important to note that the transpose of the above
expressions will not be the same, because of the
fact that $\hat{F}^T = - \hat{F}$ 
(i.e. $F_{\mu \nu} = - F_{\nu \mu}$).
Using eqn. (\ref{eqn3.20}) we can also solve for 
$x_{\mu} (s)$ and $\pi_{\mu} (s)$ in terms of 
$x_{\mu} (0)$ and $\pi_{\mu} (0)$ to get
\bea
\vec{x} (s)
&=&
\vec{x} (0) 
+ \left( \frac{e^{2 e \hat{F} s} - 1}{e \hat{F}} \right)
\cdot
\vec{\pi} (0)
+ \frac{1}{2}
\left( \frac{e^{2 e \hat{F} s} - 1}{e \hat{F}} \right)
\cdot \vec{N}
\nonumber \\
\vec{\pi} (s)
&=&
e^{2 e \hat{F} s} \cdot \vec{\pi} (0)
+ \frac{1}{2} e^{2 e \hat{F} s} \cdot \vec{N}
- \frac{1}{2} \vec{N}
\eea
Using eqn. (\ref{eqn3.30}) in eqn. (\ref{eqn3.10}) 
we get
\bea
{\cal {H}}
&=&
- \left[ x_{\mu} (s) - x_{\mu} (0) \right]
K^{\mu \nu} 
\left[ x_{\nu} (s) - x_{\nu} (0) \right]
+ m^2 + Q
\label{eqn3.40}
\eea
where
\bea
\hat{K} 
&=&
\frac{1}{4}
\left( \frac{e \hat{F}}{\sinh (e \hat{F} s)} \right)^2
\eea
Rearranging the order of the operators in 
eqn. (\ref{eqn3.40}) so that they are ordered in time
we get
\bea
{\cal {H}}
&=&
- x_{\mu} (s) K^{\mu \nu} x_{\nu} (s)
+ 2 x_{\mu} (s) K^{\mu \nu} x_{\nu} (0)
- x_{\mu} (0) K^{\mu \nu} x_{\nu} (0)
+ m^2 + Q 
- \frac{i}{2} ~\mbox{tr}~
\left[ \frac{e \hat{F}}{\tanh (e \hat{F} s)} \right]
\label{eqn3.50}
\eea
where we have used eqn. (35) to get
\bea
x_{\mu} (0) K^{\mu \nu} x_{\nu} (s) 
- x_{\mu} (s) K^{\mu \nu} x_{\nu} (0)
&=&
- \frac{i}{2} ~\mbox{tr}~
\left[ \frac{e \hat{F}}{\tanh (e \hat{F} s)} \right]
\eea
Using eqn. (\ref{eqn3.50}) and eqn. (\ref{eqn3.30})
the first order differential equations
for $U(x^{\prime\prime}, x^{\prime};s)$ in eqn. (\ref{eqn2.58}), (\ref{eqn2.59})
and (\ref{eqn2.60}) takes the form
\bea
\left\{
- (x^{\prime\prime}-x^{\prime})_{\mu}  K^{\mu \nu} (x^{\prime\prime}
-x^{\prime})_{\nu} 
+ m^2 + Q
- \frac{i}{2} ~\mbox{tr}~
\left[ \frac{e \hat{F}}{\tanh (e \hat{F} s)} \right]
\right\}
U(x^{\prime\prime},x^{\prime};s)
&=&
i \frac{\partial}{\partial s}
U(x^{\prime\prime},x^{\prime};s)
\label{eqn3.60}
\eea
\bea
\left\{
\left[ \frac{e \hat{F}}{e^{2 e \hat{F} s} - 1} 
\right]^{\mu \nu}
 (x^{\prime\prime}-x^{\prime})_{\nu} 
- \frac{1}{2} {N}^{\mu}
\right\}
U(x^{\prime\prime}, x^{\prime}; s)
&=&
\left\{
- i \frac{\partial}{\partial x^{\prime}_{\mu} }
- e A^{\mu} (x^{\prime})
\right\}
U(x^{\prime\prime}, x^{\prime};s)
\label{eqn3.61}
\\
\left\{
\left[ \frac{e \hat{F}}{1 - e^{- 2 e \hat{F} s}}
\right]^{\mu \nu}
 (x^{\prime\prime}-x^{\prime})_{\nu}  
- \frac{1}{2} {N}^{\mu}
\right\}
U(x^{\prime\prime}, x^{\prime}; s)
&=&
\left\{
+ i \frac{\partial}{\partial x_{\mu}^{\prime\prime} }
- e A^{\mu} (x^{\prime\prime})
\right\}
U(x^{\prime\prime}, x^{\prime};s)
\label{eqn3.62}
\eea
with the initial condition
in eqn. (\ref{eqn2.61}).
Integrating eqn. (\ref{eqn3.60}) with respect to $s$ on both
sides we get
\bea
U(x^{\prime\prime}, x^{\prime};s)
&=&
C(x^{\prime\prime}, x^{\prime})
\frac{1}{s^2}
e^{- i m^2 s}
e^{- i Q s}
\left[ ~\mbox{det}~ \frac{e \hat{F} s}{\sinh (e \hat{F} s)}
\right]^{\frac{1}{2}}
\hspace{3cm}
\nonumber \\
&&
\hspace{3cm}
\exp \left\{ - i \frac{1}{4 s}
	\left[ \frac{e \hat{F} s}{\tanh (e \hat{F} s)}
	\right]^{\mu \nu}
        (x^{\prime\prime}-x^{\prime})_{\mu} 
        (x^{\prime\prime}-x^{\prime})_{\nu} 
     \right\}
\label{eqn3.70} 
\eea
where $C(x^{\prime\prime}, x^{\prime})$ is the integration constant which 
has no explicit dependence on $s$.
Using the above expression for $U(x^{\prime\prime}, x^{\prime};s)$
in eqn. (\ref{eqn3.61}) and (\ref{eqn3.62})
we get the differential equations for $C(x^{\prime\prime}, x^{\prime})$ to be
\bea
\left\{
- i \frac{\partial}{\partial x_{\mu}^{\prime} }
- e A^{\mu} (x^{\prime})
+ \frac{1}{2} N^{\mu}
+ \frac{1}{2} e F^{\mu \nu} 
( x^{\prime\prime} - x^{\prime})_{\nu}  
\right\}
C(x^{\prime\prime}, x^{\prime})
&=&
0
\label{eqn3.80}
\\
\left\{
+ i \frac{\partial}{\partial x_{\mu}^{\prime\prime} }
- e A^{\mu} (x^{\prime\prime})
+ \frac{1}{2} N^{\mu}
- \frac{1}{2} e F^{\mu \nu}
(x^{\prime\prime} - x^{\prime})_{\nu}
\right\}
C(x^{\prime\prime}, x^{\prime})
&=&
0
\label{eqn3.81}
\eea
Integrating eqn. (\ref{eqn3.80}) and eqn. (\ref{eqn3.81})
we get
\bea
C(x^{\prime\prime}, x^{\prime})
&=&
C_0
~\phi (x^{\prime\prime}, x^{\prime})
\eea
where $C_0$ is the integration constant and
\bea
\phi (x^{\prime\prime}, x^{\prime})
&=&
\exp \left[
	- i e \int_{x^{\prime}}^{x^{\prime\prime}} dt_{\mu}~
	\left\{ A^{\mu} (t) + N^{\mu} \right\}
     \right]
.
\eea
$C_0$ is determined using the initial condition
in eqn. (\ref{eqn2.61}) to give
\bea
C_0 &=& - i \frac{1}{16 \pi^2}
.
\eea
where we have  used the representation for the delta function
\bea
\delta (x - x^{\prime})
&=&
\lim_{\sigma \rightarrow 0}
\frac{1}{\sqrt{2 \pi \sigma}} 
\exp \left\{ - \frac{(x - x^{\prime})^2}{2 \sigma} \right\}
.
\eea
Thus we have
\bea
U(x^{\prime\prime}, x^{\prime};s)
&=&
- i \frac{1}{16 \pi^2}
\phi (x^{\prime\prime}, x^{\prime})
\frac{1}{s^2}
e^{- i m^2 s}
e^{- i Q s}
\left[ ~\mbox{det}~ \frac{e \hat{F} s}{\sinh (e \hat{F} s)}
\right]^{\frac{1}{2}}
\hspace{3cm}
\nonumber \\
&&
\hspace{3cm}
\exp \left\{ - i \frac{1}{4 s}
	\left[ \frac{e \hat{F} s}{\tanh (e \hat{F} s)}
	\right]^{\mu \nu}
	(x^{\prime\prime}-x^{\prime})_{\mu}
	(x^{\prime\prime}-x^{\prime})_{\nu}
     \right\}
\label{eqn3.90} 
\eea
It should be emphasised that 
$\phi (x^{\prime\prime}, x^{\prime})$ does-not commute with the 
rest of the $U_0 (x^{\prime\prime}, x^{\prime};s)$.
Due to this reason it is necessary that 
$\phi (x^{\prime\prime}, x^{\prime})$ operate on to the left of $U_0 (x^{\prime\prime}, x^{\prime};s)$. 
We can evaluate $U_0 (x^{\prime\prime}, x^{\prime};s) = e^{- i {\cal {H}}_0 s}$
independently or by taking the limit $F$ and $H$ going to 
zero in eqn. (\ref{eqn3.90}) to get
\bea
U_0 (x^{\prime\prime}, x^{\prime};s)
&=&
- i \frac{1}{16 \pi^2}
\frac{1}{s^2}
e^{- i m^2 s}
{\bf 1}
\exp \left\{ - i \frac{1}{4 s}
	(x^{\prime\prime}-x^{\prime})_{\mu}
	(x^{\prime\prime}-x^{\prime})_{\nu}
     \right\}
\label{eqn3.100}
\eea
where ${\bf 1}$ is the unit matrix in the spinor space.
Using eqn. (\ref{eqn3.90}) and eqn. (\ref{eqn3.100}) we have
\bea
\mbox{Tr}~ \left[ U(x^{\prime\prime}, x^{\prime};s) \right]
&=&
- i \frac{1}{16 \pi^2}
\int d^4 x~
\frac{1}{s^2}
e^{- i m^2 s}
\left\{ ~\mbox{det}~ \frac{e \hat{F} s}{\sinh (e \hat{F} s)}
\right\}^{\frac{1}{2}}
\left\{ ~\mbox{tr}~ e^{- i Q s} \right\}
\label{eqn3.110}
\\
\mbox{Tr}~ \left[ U_0 (x^{\prime\prime}, x^{\prime};s) \right]
&=&
- i \frac{1}{16 \pi^2}
\int d^4 x~
\frac{1}{s^2}
e^{- i m^2 s}
4
\label{eqn3.111}
\eea
where 4 comes from the trace over the spinor index.
Using eqn. (\ref{eqn3.110}) and eqn. (\ref{eqn3.111}) in 
eqn. (\ref{eqn2.80}) we get
\bea
{\cal {L}}_{eff} [F, H]
&=&
- \frac{1}{4} F^2
- \frac{1}{12} H^2
+ \frac{1}{32 \pi^2}
\int_0^{\infty} \frac{ds}{s^3}
e^{- i m^2 s}
\left[ 
\left\{ ~\mbox{det}~ \frac{e \hat{F} s}{\sinh (e \hat{F} s)}
\right\}^{\frac{1}{2}}
\left\{ ~\mbox{tr}~ e^{- i Q s} \right\}
- 4
\right]
\label{eqn3.120}
\eea
Thus our problem reduces to determination of the eigenvalues
of $\hat{F}$ and $Q$. Note that $\hat{F}$ is a matrix in space time and
$Q$ is a matrix in the spinor index.
We shall determine the eigenvalues of $\hat{F}$ and $Q$
in terms of the Lorentz invariants
\bea
\begin{array}{rclcrcl}
{\cal {F}} &=& \frac{1}{4} F_{\mu \nu} F^{\mu \nu}
& \hspace{15mm} &
{\cal {G}} &=& \frac{1}{4} F_{\mu \nu} {\tilde{F}}^{\mu \nu} 
\\[2mm]
{\cal {H}} &=& \frac{1}{12} H_{\mu \nu \lambda} H^{\mu \nu \lambda}
&&
{\cal {I}}^2 &=& \frac{1}{48} (H_{\lambda \mu \nu} F^{\mu \nu})
(H^{\lambda \alpha \beta} F_{\alpha \beta})
= - \frac{1}{12} ({\tilde{H}}_{\mu} {\tilde{F}}^{\mu \lambda})
({\tilde{F}}_{\lambda \nu} {\tilde{H}}^{\nu})
\end{array}
\eea
where ${\tilde{F}}^{\mu \nu}$ is the dual tensor corresponding to
$F^{\mu \nu}$ and is given by
${\tilde{F}}^{\mu \nu} = \frac{1}{2}
\epsilon^{\mu \nu \alpha \beta} F_{\alpha \beta}$.

Eigen values of $\hat{F}$ is determined \cite{sch1951} by observing
that $\hat{F}$ satisfies the identity
\bea
\hat{F}^4 + 2 {\cal {F}} \hat{F}^2 - {\cal {G}}^2 = 0
\label{eqn3.130}
\eea
This can be verified using the identities
\be
F_{\mu \lambda} {\tilde{F}}^{\lambda \nu}
=
- {\delta_{\mu}}^{\nu}
{\cal {G}}
\hspace{10mm}
\mbox{and}
\hspace{10mm}
{\tilde{F}}_{\mu \lambda} {\tilde{F}}^{\lambda \nu}
- F_{\mu \lambda} F^{\lambda \nu}
= 
2 {\delta_{\mu}}^{\nu} {\cal {F}}
.
\ee
Using eqn. (\ref{eqn3.130}) we have the eigenvalues of
$\hat{F}$ to be
$\pm \lambda_1$ and $\pm \lambda_2$
where
\bea
\lambda_1
&=&
\frac{i}{\sqrt{2}}
\left[ 
\left\{ {\cal {F}} + i {\cal {G}} \right\}^{\frac{1}{2}}
+
\left\{ {\cal {F}} - i {\cal {G}} \right\}^{\frac{1}{2}}
\right]
\\
\lambda_2
&=&
\frac{i}{\sqrt{2}}
\left[
\left\{ {\cal {F}} + i {\cal {G}} \right\}^{\frac{1}{2}}
-
\left\{ {\cal {F}} - i {\cal {G}} \right\}^{\frac{1}{2}}
\right]
\label{eqn3.140}
\eea
Thus we have
\bea
\left\{
~\mbox{det}~
\frac{e \hat{F} s}{\sinh (e \hat{F} s)}
\right\}^{\frac{1}{2}}
&=&
\frac{(e \lambda_1 s)}{\sinh (e \lambda_1 s)}
\frac{(e \lambda_2 s)}{\sinh (e \lambda_2 s)}
\label{eqn3.143}
\eea

We shall now determine the eigenvalues of $Q$.
Beginning from the expression for $Q$ in eqn. (\ref{eqn3.11})
we have
\bea
Q^2 
&=&
(a - d) {\rm 1} + i c_{\alpha} \gamma^{\alpha}
+ i b \gamma_5
\label{eqn3.150}
\eea
and
\bea
Q^4
&=&
\left\{ (a -d)^2 - b^2 - c^2 \right\} {\rm 1}
+ 2 i (a -d) c_{\alpha} \gamma^{\alpha}
+ 2 i (a-d) b \gamma_5
\label{eqn3.160}
\eea
where
\bea
\begin{array}{rclcrcl}
a &=& 2 e^2 {\cal {F}} & \hspace{15mm} & b &=& 2 e^2 {\cal {G}} \\
d &=& 2 g^2 {\cal {H}} &  & 
c_{\alpha} &=& 4 \sqrt{3} e g {\cal {I}}_{\alpha}
\end{array}
\eea
In arriving at eqn. (\ref{eqn3.150}) and eqn. (\ref{eqn3.160})
we have used the identities
\be
\left\{ 
\sigma_{\mu \nu} , \sigma_{\alpha \beta}
\right\}
=
2 ( g_{\mu \alpha} g_{\nu \beta}
	- g_{\mu \beta} g_{\nu \alpha})
{\rm 1}
+
2 i \epsilon_{\mu \nu \alpha \beta} \gamma_5
\hspace{10mm}
\mbox{and}
\hspace{10mm}
\left\{
\sigma_{\mu \nu} , \gamma_5 \gamma_{\alpha}
\right\}
=
- 2 \epsilon_{\mu \nu \alpha \sigma} \gamma^{\sigma}
\ee
and
\be
{\tilde{H}}_{\mu} {\tilde{H}}^{\mu}
=
- \frac{1}{6}
H_{\mu \nu \lambda} H^{\mu \nu \lambda}
\hspace{10mm}
\mbox{and}
\hspace{10mm}
{\tilde{H}}^{\mu} {\tilde{F}}_{\mu \lambda}
=
- \frac{1}{2} H_{\lambda \alpha \beta} F^{\alpha \beta}
\ee
Using eqn. (\ref{eqn3.150}) and eqn. (\ref{eqn3.160})
it is easy to verify that $Q$ satisfies the relation
\bea
Q^4 - 2 (a - d) Q^2 
+ \left\{ (a -d)^2 + b^2 + c^2 \right\}
&=&
0
\eea
Thus the eigenvalues of $Q$ are $\pm e q_1$
and $\pm e q_2$, where
\bea
e q_1
&=&
\sqrt{2}
\left\{
(e^2 {\cal {F}} - g^2 {\cal {H}})
+ i 
\sqrt{e^4 {\cal {G}}^2 + 12 e^2 g^2 {\cal {I}}^2}
\right\}^{\frac{1}{2}}
\nonumber 
\\
e q_2
&=&
\sqrt{2}
\left\{
(e^2 {\cal {F}} - g^2 {\cal {H}})
- i
\sqrt{e^4 {\cal {G}}^2 + 12 e^2 g^2 {\cal {I}}^2}
\right\}^{\frac{1}{2}}
\label{eqn3.170}
\eea
Using eqn. (\ref{eqn3.170}) we have
\bea
\left\{
~\mbox{tr}~ e^{- i Q s}
\right\}
&=&
4 \cosh (e \chi_1 s) \cosh (e \chi_2 s)
\label{eqn3.180}
\eea
where
\bea
e \chi_1
&=&
\frac{i}{\sqrt{2}}
\left[
\left\{
(e^2 {\cal {F}} - g^2 {\cal {H}})
+ i
\sqrt{e^4 {\cal {G}}^2 + 12 e^2 g^2 {\cal {I}}^2}
\right\}^{\frac{1}{2}}
+
\left\{
(e^2 {\cal {F}} - g^2 {\cal {H}})
- i
\sqrt{e^4 {\cal {G}}^2 + 12 e^2 g^2 {\cal {I}}^2}
\right\}^{\frac{1}{2}}
\right]
\nonumber 
\\
e \chi_2
&=&
\frac{i}{\sqrt{2}}
\left[
\left\{
(e^2 {\cal {F}} - g^2 {\cal {H}})
+ i
\sqrt{e^4 {\cal {G}}^2 + 12 e^2 g^2 {\cal {I}}^2}
\right\}^{\frac{1}{2}}
-
\left\{
(e^2 {\cal {F}} - g^2 {\cal {H}})
- i
\sqrt{e^4 {\cal {G}}^2 + 12 e^2 g^2 {\cal {I}}^2}
\right\}^{\frac{1}{2}}
\right]
\eea
Observe that
in the limit
$g \rightarrow 0$ we have $\chi_1 \rightarrow \lambda_1$
and $\chi_2 \rightarrow \lambda_2$.
Using eqn. (\ref{eqn3.143}) and eqn. (\ref{eqn3.180})
in eqn. (\ref{eqn3.120}) we have
\bea
{\cal {L}}_{eff} [F, H]
&=&
- {\cal {F}} - {\cal {H}}
+ \frac{1}{8 \pi^2}
\int_0^{\infty} \frac{ds}{s^3}
e^{- i m^2 s}
\left[ 
(e \lambda_1 s) (e \lambda_2 s)
\frac{\cosh (e \chi_1 s)}{\sinh (e \lambda_1 s)}
\frac{\cosh (e \chi_2 s)}{\sinh (e \lambda_2 s)}
- 1
\right]
\eea
Expressing the above integral in terms of the contour running 
along the positive real axis 
(cf. eqn. (\ref{eqn2.50}) )
which is effectively got by substituting 
$s \rightarrow -is$ in the integral,
we get
\bea
{\cal {L}}_{eff} [F, H]
&=&
- {\cal {F}} - {\cal {H}}
- \frac{1}{8 \pi^2}
\int_0^{\infty} \frac{ds}{s^3}
e^{- m^2 s}
\left[ 
(e \lambda_1 s) (e \lambda_2 s)
\frac{\cos (e \chi_1 s)}{\sin (e \lambda_1 s)}
\frac{\cos (e \chi_2 s)}{\sin (e \lambda_2 s)}
- 1
\right]
\eea

It should be noted that $s\rightarrow is$ is permissible provided there are no
singularities associated with pair creation. Separating the divergent terms in 
the above expression
we have
\bea
{\cal {L}}_{eff} [F, H]
&=&
- (1 + e^2 C_e) {\cal {F}} 
- (1 - g^2 C_g) {\cal {H}}
\hspace{40mm}
\nonumber 
\\
&&
\hspace{10mm}
- \frac{1}{8 \pi^2}
\int_0^{\infty} \frac{ds}{s^3}
e^{- m^2 s}
\left[ 
(e \lambda_1 s) (e \lambda_2 s)
\frac{\cos (e \chi_1 s)}{\sin (e \lambda_1 s)}
\frac{\cos (e \chi_2 s)}{\sin (e \lambda_2 s)}
- 1
- \frac{2}{3} e^2 s^2 {\cal {F}}
+ g^2 s^2 {\cal {H}}
\right]
\eea
where
\be
C_e = \frac{1}{12 \pi^2} \int_0^{\infty} \frac{ds}{s} e^{- m^2 s}
\hspace{10mm}
\mbox{and}
\hspace{10mm}
C_g = \frac{1}{8 \pi^2} \int_0^{\infty} \frac{ds}{s} e^{- m^2 s}
.
\ee
We can absorb the divergences $C_e$ and $C_g$ by suitable
change of scale of the fields and couplings.
Identifying the quantities used till now by the subscript
zero, we define the following scale transformations
\bea
\begin{array}{rclcrclcrcl}
e^2 &=& \frac{e_0^2}{1 + e_0^2 C_e}
&\hspace{10mm}&
{\cal {F}} &=& (1 + e_0^2 C_e) {\cal {F}}_0  
&\hspace{10mm}&
{\cal {G}} &=& (1 + e_0^2 C_e) {\cal {G}}_0
\\
g^2 &=& \frac{g_0^2}{1 - g_0^2 C_g}
&&
{\cal {H}} &=& (1 - g_0^2 C_g) {\cal {H}}_0
&&
{\cal {I}}^2 &=& (1 + e_0^2 C_e) (1 - g_0^2 C_g) {\cal {I}}_0^2
\end{array}
\label{eqn3.190}
\eea
Using the above scale transformations we obtain the divergence free
effective Lagrangian to be
\bea
{\cal {L}}_{eff} [F, H]
&=&
- {\cal {F}} - {\cal {H}}
- \frac{1}{8 \pi^2}
\int_0^{\infty} \frac{ds}{s^3}
e^{- m^2 s}
\left[ 
(e \lambda_1 s) (e \lambda_2 s)
\frac{\cos (e \chi_1 s)}{\sin (e \lambda_1 s)}
\frac{\cos (e \chi_2 s)}{\sin (e \lambda_2 s)}
- 1
- \frac{2}{3} e^2 s^2 {\cal {F}}
+ g^2 s^2 {\cal {H}}
\right]
\eea
This is the generalisation of the Euler-Heisenberg lagrangian to include constant
Maxwell and Kalb-Ramond fields. Explicitly expanding the above expression for weak 
fields using the series expansions for $\cos x$ and $(\sin x)^{-1}$ we get
\bea
{\cal {L}}_{eff} [F, H]
&=&
- {\cal {F}} - {\cal {H}}
+ \frac{1}{8 \pi^2} \frac{1}{m^4}
\left\{
\frac{4}{45}~ e^4 {\cal {F}}^2
+ \frac{7}{45}~ e^4 {\cal {G}}^2
- \frac{1}{6}~ g^4 {\cal {H}}^2
+ 2~ e^2 g^2 {\cal {I}}^2
\right\}
+ ...
\eea
In addition to the self-interaction term $-{1\over
6}g^4{\cal{H}}^2$ for the Kalb-Ramond fields an interaction term
between the Maxwell field 
and the Kalb-Ramond field given by $2e^2g^2{\cal {I}}^2$ is also
induced in the leading order.
Observe that for the case of $g \rightarrow 0$ we get back 
Schwinger's result. It is interesting to note that the sign of the
${\cal {H}}^2$ term is negative.


\section{Acknowledgements}

K. V. Shajesh would like to express his appreciation
to The Institute 
of Mathematical Sciences for his visit during which this work was undertaken.



\end{document}